\documentclass[prd,a4paper,twocolumn, superscriptaddress,floatfix,nofootinbib]{revtex4}
\usepackage{graphicx}
\usepackage{bbm}
\usepackage{amssymb}
\usepackage{hyperref}

\newcommand{\be}{\begin{equation}}
\newcommand{\ee}{\end{equation}}
\newcommand{\bq}{\begin{eqnarray}}
\newcommand{\eq}{\end{eqnarray}}
\newcommand{\bsq}{\begin{subequations}}
\newcommand{\esq}{\end{subequations}}
\newcommand{\bc}{\begin{center}}
\newcommand{\ec}{\end{center}}
\newcommand\lapp{\mathrel{\rlap{\lower4pt\hbox{\hskip1pt$\sim$}} \raise1pt\hbox{$<$}}}
\newcommand\gapp{\mathrel{\rlap{\lower4pt\hbox{\hskip1pt$\sim$}} \raise1pt\hbox{$>$}}}

\newcommand{\dtot}[2]{\frac{d #1}{d #2}}

\newcommand{\vv}{{\bar v}}
\newcommand{\cc}{{\tilde c}}
\begin{document}

\title{Revisiting the velocity-dependent one-scale model for monopoles}

\author{L. Sousa}
\email[Electronic address: ]{Lara.Sousa@astro.up.pt}
\affiliation{Instituto de Astrof\'{\i}sica e Ci\^encias do Espa{\c c}o, Universidade do Porto, CAUP, Rua das Estrelas, PT4150-762 Porto, Portugal}
\affiliation{Centro de Astrof\'{\i}sica da Universidade do Porto, Rua das Estrelas, PT4150-762 Porto, Portugal}

\author{P.P. Avelino}
\email[Electronic address: ]{pedro.avelino@astro.up.pt}
\affiliation{Instituto de Astrof\'{\i}sica e Ci\^encias do Espa{\c c}o, Universidade do Porto, CAUP, Rua das Estrelas, PT4150-762 Porto, Portugal}
\affiliation{Centro de Astrof\'{\i}sica da Universidade do Porto, Rua das Estrelas, PT4150-762 Porto, Portugal}
\affiliation{Departamento de F\'{\i}sica e Astronomia, Faculdade de Ci\^encias, Universidade do Porto, Rua do Campo Alegre 687, PT4169-007 Porto, Portugal}

\begin{abstract}
We revisit the physical properties of global and local monopoles and discuss their implications in the dynamics of monopole networks. In particular, we review the Velocity-dependent One-Scale (VOS) model for global and local monopoles and propose physically motivated changes to its equations. We suggest a new form for the acceleration term of the evolution equation of the root-mean-squared velocity and show that, with this change, the VOS model is able to describe the results of radiation and matter era numerical simulations of global monopole networks with a single value of the acceleration parameter $k$, thus resolving the tension previously found in the literature. We also show that the fact that the energy of global monopoles is not localized within their cores affects their dynamics and, thus, the Hubble damping terms in the VOS equations. We study the ultra-relativistic linear scaling regime predicted by the VOS equations and demonstrate that it cannot be attained either on radiation or matter eras and, thus, cannot arise from the cosmological evolution of a global monopole network. We also briefly discuss the implications of our findings for the VOS model for local monopoles.
\end{abstract}

\maketitle
\flushbottom

\date{\today}


\section{Introduction}

The production of topological defects in symmetry-breaking phase transitions in the early universe is expected in a large variety of Grand Unified theories \cite{vilenkin2000cosmic}. The dimensionality of the defects that are created is determined by the type of symmetry that is broken: monopoles may be created when there is a breaking of spherical symmetry; line-like defects known as cosmic strings may be formed when an axial symmetry is broken; and $2+1$-dimensional defects dubbed domain walls may be formed when a discrete symmetry is broken. Although the production of these defects may occur in the early universe, they are expected to survive throughout cosmological history, potentially leaving behind distinct signatures on a variety of observational probes. Describing the evolution of topological defect networks is necessary in order to accurately characterize these observational signatures. Although this can be done using numerical simulations, these may be computationally costly and limited in dynamical range. One may also resort to analytical models to describe the evolution of topological defects, which --- when calibrated with the aid of simulations --- are often more versatile, allowing for accurate predictions of the observational signatures of cosmic defects.

Although the properties of topological defect networks are dependent on the defects' codimension, their macroscopic dynamics may be described in a unified framework by resorting to a semi-analytical model known as Velocity-dependent One-Scale (VOS) model. This model --- which may be derived from the (generalized) Nambu-Goto action --- describes the cosmological evolution of networks of defects of arbitrary dimensionality by following the evolution of two variables: the characteristic lenghtscale $L$ and its Root-Mean-Squared (RMS) velocity $\vv$ \cite{Sousa:2011ew,Sousa:2011iu}. This characteristic lengthscale is a measure of the energy density of defects and is defined as

\be
\rho=\frac{\sigma_p}{L^{3-p}}\label{denene}\,,
\ee
where $\rho$ is the average topological defect energy density, $p$ is the dimensionality of the defect (with $p=0,1,\,\mbox{and}\,\,2$ for point particles, strings and domain walls in $3+1$-dimensions respectively), and $\sigma_p$ is the defect mass per unit $p$-dimensional area (where $\sigma_0\equiv M$ is the particle mass, $\sigma_1\equiv\mu$ is the cosmic string tension, and $\sigma_2\equiv\sigma$ is the surface tension of a domain wall). In this case, the evolution equations for $\vv$ and $L$ are of the form

\bq
\dtot{\vv}{t} & = & \left(1-\vv^2\right)\left[\frac{\kappa}{L}-\frac{\vv}{\ell_d}\right]\,,\label{vosv}\\
\dtot{L}{t} & = & HL+\frac{1}{D}\frac{L}{\ell_d}\vv^2+\frac{\cc}{D}\vv\,, \label{vosL}
\eq
where $H=(da/dt)/a$ is the Hubble parameter, $a$ is the cosmological scale factor, and $D=3-p$. Here, we have also introduced the damping lengthscale $\ell_d^{-1}=\left(p+1\right)H+\ell_f^{-1}$ that includes not only the effects of cosmological expansion, but also of the frictional forces resulting from particle scattering (encoded in the frictional lengthscale $\ell_f$). The last term in Eq. (\ref{vosL}) is a phenomenological term that accounts for the energy loss caused by defect interactions, $\cc$ is a phenomenological parameter that quantifies the efficiency of this energy loss mechanism, and $\kappa$ is an adimensional momentum parameter. This parameter characterizes the acceleration felt by the topological defects. In the case of cosmic strings or domain walls, this acceleration is mainly caused by their linear or surface tension respectively. This source of acceleration is not present in the case of minimally interacting point particles for which $\kappa= 0$. Note however that, for monopoles that interact non-minimally with each other, the specific form of this acceleration term will depend on the type of interaction.

The VOS model provides a good description of the evolution of topological defect networks on sufficiently large scales (see \cite{Avelino:2015kdn} for a derivation of these equations from thermodynamical principles). In particular, it has been shown to provide an accurate description of the cosmological evolution of cosmic string \cite{Martins:1996jp} and domain wall \cite{Avelino:2005kn} networks from early to late cosmological times, with a unique calibration of the parameters $\kappa$ and $\cc$. However, one would need to adapt this model in order to be able to use it to describe the dynamics of realistic monopole networks.

Strictly speaking, with $p=0$, these equations describe the evolution of a network of minimally interacting point particles. Although they capture essential aspects of monopoles dynamics, both global and local monopoles have specific properties that are not taken into account in this model. A first attempt to describe realistic monopole networks was done in \cite{Martins:2008zz}. However, subsequent numerical simulations of global monopoles networks \cite{Lopez-Eiguren:2016jsy} have shown that the model introduced in \cite{Martins:2008zz} fails to describe the macroscopic dynamics of these networks with a single value of the acceleration parameter. Here, we revisit the problem of extending the VOS model to describe global and local monopoles, and propose physically motivated changes to this model which allow for an improved description of the dynamics of monopole networks in a cosmological setting.

This paper is organized as follows. In Sec. \ref{global}, we revisit the properties of global monopoles and propose corresponding alterations to the VOS equations that describe the cosmological dynamics of these networks. In Subsec. \ref{linear}, we discuss the linear scaling regime and use the results of the most accurate numerical simulations to estimate the values of the VOS parameters. In Subsec. \ref{ultra}, we discuss the ultra-relativistic linear scaling regime and argue that this regime cannot be the end result of the evolution of global monopole networks. In Sec. \ref{local}, we briefly discuss the properties of local monopoles and propose changes to the VOS model to account for their specific properties. We also briefly discuss the scaling regimes in this model. We then conclude in Sec. \ref{conc}.

\section{Global Monopoles}\label{global}

The energy of global monopoles is not localized within their cores and, thus, treating them as point particles is inadequate. As a matter of fact, the total energy of a global monopole grows linearly with distance \cite{vilenkin2000cosmic,Barriola:1989hx} and, as a consequence, they exert long-range forces on each other. It is straightforward to show that these forces are independent of distance and have a magnitude \cite{vilenkin2000cosmic,Barriola:1989hx}

\be
F\sim4\pi \eta^2\,,
\label{forceglobal}
\ee
where $\eta$ is the energy scale of the monopole-creating phase transition (note that the total mass of a global monopole at a distance $L$ is $M\sim4\pi \eta^2 L$). This force, as pointed out in \cite{Martins:2008zz}, gives rise to an acceleration of the form $k/L$ (where $k$ is a constant). Note, however, that a monopole or anti-monopole feels the acceleration caused by each monopole and anti-monopole within its causal volume. Since each of the monopoles and anti-monopoles exerts a force of similar magnitude, the problem of computing the total acceleration acting on a monopole is then analogous to finding the distance traveled in a $3$-dimensional random walk with constant step. If the average number of monopoles (and anti-monopoles) per cosmic horizon ---

\be
\mathcal{N}=\left(\frac{d_H}{L}\right)^3\label{number}\,,
\ee
where $d_H$ is the cosmological horizon --- is large and the positions of monopoles/anti-monopoles are uncorrelated, we expect the total acceleration to be approximately

\be
\sqrt{\mathcal{N}}\frac{k}{L}\label{curv1}\,,
\ee
since, in this case, the average number of monopoles and anti-monopoles within the causal volume of any given monopole will also be given by $\mathcal{N}$. Here, $k$ is an adimensional acceleration parameter, which should not depend strongly on the cosmological background. Note, however, that numerical simulations \cite{Bennett:1990xy,Yamaguchi:2001xn,Lopez-Eiguren:2016jsy} seem to indicate that $\mathcal{N}$ is small ($\mathcal{N}\sim 2-7$ was measured), and, thus, the average total force acting on a monopole may deviate from Eq. (\ref{curv1}). In order to account for this possibility, we introduce the following parameterization for the acceleration term

\be
\frac{k}{L}\left(\frac{d_H}{L}\right)^{\alpha}\label{curvterm}\,.
\ee
One would expect $\alpha\sim0$ when there is, on average, only one other monopole or anti-monopole per cosmological horizon, and $\alpha \to 3/2$ for large $\mathcal{N}$. So one should expect $0\le \alpha\le 3/2$. Negative values of $\alpha$ --- such as the value $\alpha=-3/2$ suggested in the VOS model for global monopoles in Ref. \cite{Martins:2008zz} --- are not to be expected except perhaps for very fine-tuned (and thus unrealistic) configurations of monopoles and anti-monopoles. The parameterization in Eq. (\ref{curvterm}) will allow us to investigate the validity of these expectations and to compare the adequacy of these two choices for $\alpha$ ($\alpha=-3/2$ and $\alpha=3/2$).

Moreover, for global monopoles, the main energy loss mechanism is the annihilation of monopole and anti-monopoles pairs (which carry opposite topological charges). Monopoles and anti-monopoles are attracted to each other and may evolve to create bound states. Once a monopole-anti-monopole bound pair is created, they move at ultra-relativistic speeds and loose energy by emitting Goldstone-boson radiation. As a consequence, the separation between the monopole and anti-monopole decreases until they eventually annihilate within a Hubble time \cite{Barriola:1989hx}. It is this process of monopole-anti-monopole annihilation that provides an efficient energy loss mechanism and explains the small values of $\mathcal{N}$ measured in numerical simulations. Note however that the existence of such processes also hints that the evolution of a global monopole network may be misrepresented by a simple two-parameter VOS model. 

The existence of long range forces between monopoles/anti-monopoles has other consequences beyond the form of the acceleration term. For global monopoles, the average energy density is of the form

\be
\rho=\frac{M}{L^3}=\frac{4\pi\eta^2 L}{L^3}=\frac{4\pi \eta^2}{L^2}\label{rhonew}\,,
\ee
which --- similarly to that of cosmic strings and unlike point particles --- scales as $L^{-2}$. This is a consequence of the fact that the mass of global monopoles grows linearly with distance, which decreases the dependence of $\rho$ on $L$. This fact will necessarily affect the coefficients of the Hubble damping terms in the VOS equations (see Eqs. (\ref{vosv}) and (\ref{vosL})). See Ref. \cite{Avelino:2015kdn} for a discussion of the dynamical effects associated to variations of the defect mass per unit p-dimensional area, $\sigma_p$.

Given these properties of global monopoles, we propose the following equations for $\vv$ and $L$:

\bq
\dtot{\vv}{t} & = & \left(1-\vv^2\right)\left[\frac{k}{L}\left(\frac{d_H}{L}\right)^{\alpha} - \frac{\vv}{\ell_d}\right]\label{vosvnew}\,,\\
\dtot{L}{t} & = & HL +\frac{1}{\theta}\frac{L}{\ell_d}\vv^2+\frac{\cc}{\theta}\vv\label{vosLnew}\,,
\eq
where $\ell_d^{-1}=\lambda H+\ell_f^{-1}$, and one would expect $\theta=\lambda=2$ (since $\rho \propto L^{-2}$). These equations provide a VOS model for global monopoles with three free parameters $(k,\cc,\alpha)$, which differs from the one in \cite{Martins:2008zz} in the form of the acceleration term and in the damping coefficients (which were assumed to be $\theta=3$ and $\lambda=1$ originally). In the next subsection we shall demonstrate that this new VOS model is able to describe the results of radiation and matter era numerical simulations of global monopole networks with a single value of the acceleration parameter $k$. In order to allow for a comparison with the original model, we shall consider two situations: $\theta=3$ and $\lambda=1$ --- which has the underlying assumption that the damping effect caused by the expansion is analogous to that felt by point particles --- and $\theta=\lambda=2$ --- which treats monopoles as rigid bodies whose mass increases proportionally to the characteristic length $L$. Although reality is likely significantly more complex than either of these situations, the parameter choica $\theta=\lambda=2$ is better motivated from the physical point of view. If monopoles do not behave as rigid bodies, the realistic value of $\lambda$ may differ from $\lambda=2$ since this value follows from that assumption (as to the value of $\theta=2$ it follows simply from Eq. (\ref{rhonew}) and thus one should expect it to hold). The determination of the set of parameters of our model that best describes the dynamics of global monopoles would give us more detailed information about the properties of global monopoles themselves. It is thus a rather interesting question that warrants further investigation in numerical simulations.

\subsection{Linear Scaling regime and parameter fitting}\label{linear}

Numerical simulations of global monopole networks \cite{Bennett:1990xy,Yamaguchi:2001xn,Lopez-Eiguren:2016jsy} have demonstrated that they evolve towards a linear scaling regime during which

\be
L=\xi t \,,\qquad\mbox{and}\qquad \vv=\mbox{constant}\,,
\ee
with constant $\xi$, both in the matter- and radiation-dominated epochs. Eqs. (\ref{vosvnew}) and (\ref{vosLnew}) admit solutions of this form in the case of a power-law cosmological expansion, with $a\propto t^\beta$, in the frictionless regime (with $\ell_f=+\infty$). This regime would be characterized by

\be
\xi=\frac{\cc \vv}{\theta\left(1-\beta\right)-\beta\lambda\vv^2}\,,\quad\mbox{and}\quad\vv=\frac{k}{\lambda \beta \left(1-\beta\right)^{\alpha}\xi^{\alpha+1}}\,.
\ee
The VOS equations (Eqs. (\ref{vosvnew}) and (\ref{vosLnew})) may also admit an ultra-relativistic linear scaling solution (with $\vv=1$) that will be discussed in the next subsection.

Since numerical simulations have established the existence of a (subluminal) linear scaling regime, the measured values of $\xi$ and $\vv$ may be used to calibrate the values of $k$ and $\cc$:

\bq
k & = & \lambda \vv \beta \left(1-\beta\right)^{\alpha} \xi^{\alpha+1}\,,\label{kpar}\\
\cc & = & \frac{\xi}{\vv}\left[\theta \left(1-\beta \right)-\beta \lambda \vv^2 \right]\label{cpar}\,.
\eq
The authors of \cite{Lopez-Eiguren:2016jsy} did precisely that using the VOS model for global monopoles developed in \cite{Martins:2008zz} (characterized by $\alpha=-3/2$, $\theta=3$, and $\lambda=1$). The (averaged) values measured for the scaling parameters in the radiation era,

\be
\xi_r=1.47\pm0.09\,,\qquad\mbox{and}\qquad v_r=0.76\pm0.07\,,\label{radpar}
\ee
and in the matter era,

\be
\xi_m=1.98\pm0.07\,,\qquad\mbox{and}\qquad v_m=0.65\pm0.08\,,\label{matpar}
\ee
yielded, in both eras, values of $\cc$ that are compatible\footnote{However, there is an error in the expression for $\cc$ used in Ref. \cite{Lopez-Eiguren:2016jsy} (in Eq. (5.4)), and, for this reason, the values of the energy loss parameter obtained therein (as well as the uncertainties) differ from the ones of the present paper.}. However, a tension between the inferred values of $k$ in the matter and radiation eras was found in \cite{Lopez-Eiguren:2016jsy}. Nevertheless, although the expression of $\cc$ is independent of $\alpha$, that is not the case for $k$, and thus our physically motivated changes to the form of the acceleration term may resolve this tension.

\begin{figure}
\centering
\includegraphics[width=3.4in]{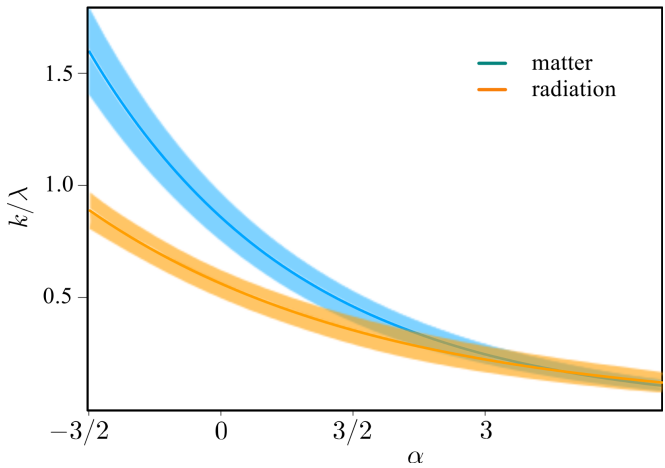}
\caption{The value of the acceleration parameter $k$ measured using the radiation (orange line) and matter (blue line) era simulations in \cite{Lopez-Eiguren:2016jsy} as a function of $\alpha$. The shaded areas represent the uncertainties associated with the determination of $k$ (obtained using linear propagation of the uncertainties in $\xi$ and $\vv$).}
\label{curvature}
\end{figure}

To study this possibility, we have computed the values of $k$ using the values of $\xi$ and $\vv$ obtained in radiation and matter era simulations (given in Eqs. (\ref{radpar}) and (\ref{matpar})), as a function of $\alpha$. The results are plotted in Fig. \ref{curvature}, as well as the errors associated to this computation, which were calculated using a linear propagation of uncertainties. Fig. \ref{curvature} shows that, independently of the choice of $\theta$ and $\lambda$, the tension between the values of $k$ in the radiation and matter eras is resolved if $\alpha$ is sufficiently large. In particular, for $\alpha \gtrsim 1.15$, the values of the parameter $k$ obtained for the matter and radiation eras seem to be compatible with each other within error margins. Our proposed value of $\alpha=3/2$ is well within the allowed interval of $\alpha$. In this case ($\alpha=3/2$), for $\theta=\lambda=2$, the values of the parameter $k$ for the radiation- and matter-dominated epochs are

\be
k_r=0.7\pm0.13\qquad\mbox{and}\qquad k_m=0.92\pm0.14\,,
\ee
which are not only compatible but also appear to be perfectly reasonable values for this parameter. For $\theta=3$ and $\lambda=1$, one obtains $k_r=0.35\pm0.06$ and $k_m=0.46\pm0.07$. Note, however, that the values of $k_r$ and $k_m$ can be reconciled for a large range of (positive) values of $\alpha$, since $k$ varies rather slowly as a function of $\alpha$ in this range. As matter of fact, $k_m=k_r$ only happens when $\alpha\simeq 4$. Still, as we have discussed, one would not expect $\alpha$ to be larger than $3/2$ (or, being less restrictive, $3$ --- which would correspond to the rather unrealistic situation in which all monopoles are perfectly aligned). To shorten this range, one would either need to decrease the systematic error of numerical simulations or to run simulations with other values of $\beta$ for calibration.

As to the parameter $\cc$, which depends on the coefficients $\theta$ and $\lambda$, the results for the radiation and matter eras are compatible in both cases under study. For $\theta=\lambda=2$, these yield, respectively,

\be
c_r=0.82\pm0.29\qquad\mbox{and}\qquad c_m=0.31\pm0.46\,,
\ee
where the uncertainties were also computed using linear propagation of the errors in $\xi$ and $\vv$. These results are compatible, but the uncertainties are rather larger (despite the error in $\vv$ and $\xi$ being small) as a consequence of the non-linear dependence of $\cc$ on $\vv$. Note, however, that given Eq. (\ref{cpar}), a significant reduction of the magnitude of this uncertainty would require a significant reduction in the error in the determination of $\xi$ and $\vv$ (particularly, in $\vv$). As for the values of $\cc$ for $\theta=3$ and $\lambda=1$, we found

\be
c_r=2.34\pm0.35\qquad\mbox{and}\qquad c_m=2.19\pm0.49.
\ee
These results are also compatible. Although the uncertainties are of the same magnitude, they result in smaller relative errors because the predicted values of $\cc$ are larger. The values of $\cc$, however, may seem atypical: for both cosmic strings and domain walls $\cc$ is smaller than unity. Nevertheless, we must stress that the scenario in which $\theta=\lambda=2$ appears to be more physically motivated and its what one would (naively) expect given the fact that energy is not localized within the monopoles' cores.

These results show that the physically motivated changes to the VOS equations we have proposed allow for an improved description of numerical simulations of global monopole networks. As a matter of fact, the proposed changes to the acceleration term of the evolution equation for $\vv$ (particularly, having $\alpha\ge 1.15$) are essential to describe both matter and radiation era simulations with a unique value of the acceleration parameter $k$.

\subsection{The ultra-relativistic regime}\label{ultra}

Eqs. (\ref{vosvnew}) and (\ref{vosLnew}) also admit an ultra-relativistic linear scaling solution characterized by

\be
\vv=1\qquad\mbox{and}\qquad\xi_s=\frac{\cc}{\theta(1-\beta)-\beta\lambda}\label{scalultra}\,,
\ee
for $\beta<\theta/(\theta+\lambda)$. This means that, for $\theta=\lambda=2$, this ultra-relativistic regime would only be allowed for $\beta<1/2$ and this fact could explain why this regime was not observed in the radiation and matter era numerical simulations of \cite{Lopez-Eiguren:2016jsy}. For $\theta=3$ and $\lambda=1$, this regime is admissible, in principle, for $\beta<3/4$ and, thus, in both matter and radiation eras.

\begin{figure}
\centering
\includegraphics[width=3.3in]{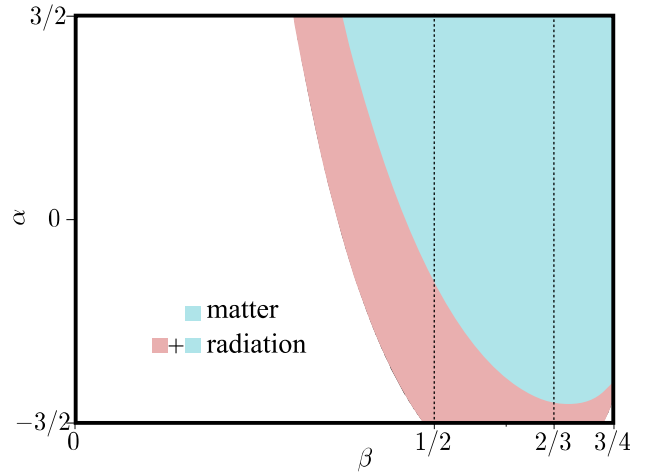}
\caption{Region of parameter space $(\beta,\alpha)$ for which the ultra-relativistic linear scaling regime is unstable, assuming that $\theta=3$ and $\lambda=1$. The blue shaded area corresponds to the region which is excluded using the values of $k$ and $\cc$ inferred from matter era simulations for calibration, while the pink shaded area corresponds to the additional exclusion region obtained when radiation era simulations are used for calibration.}
\label{ultrarelativistic}
\end{figure}

Note however that this is not the only restriction that applies to this regime and that the range of values of $\beta$ for which it is attainable may be considerably smaller. Although the regime in Eq. (\ref{scalultra}) is an equilibrium point of the VOS equations, it will only result from the evolution of the monopole network if it is a stable attractor. Let us assume that the network is initially in the regime defined in Eq. (\ref{scalultra}) and that $\vv$ and $\xi$ are perturbed such that $\vv=1-\delta v$ and $\xi=\xi_s+\delta \xi$. We then have, to first order in $\delta v$ and $\delta \xi$, that

\be
\dtot{(\delta v)}{t}=-\frac{2\delta v}{t}\left\{\frac{k}{\cc^{\alpha+1}}\frac{\left[\theta(1-\beta)-\beta\lambda\right]^{\alpha+1}}{(1-\beta)^{\alpha}}-\lambda \beta\right\}\,.
\ee
Therefore, this ultra-relativistic scaling regime is only attainable if the quantity in brackets is positive. The range of $\beta$ for which this regime is stable is dependent on $\alpha$ and on the parameters $k$ and $\cc$. However, since we shall investigate it for different values of $\alpha$, no unique calibration of $k$ and $\cc$ exists (particularly for negative values of $\alpha$, since there is a tension between the values of $k_m$ and $k_r$ that result from simulations). For this reason, we chose to study the stability using the two different calibrations that result from radiation and matter simulations. In particular, when using radiation (matter) era simulations for calibration, we use Eqs. (\ref{kpar}) and (\ref{cpar}) to compute, for each value of $\alpha$ and $\beta$, the values of $k$ and $\cc$ using the central values of the scaling parameters in Eq. (\ref{radpar}) (Eq. (\ref{matpar})). Since, for $\theta=\lambda=2$, it is clear that the ultra-relativistic regime is not allowed both in the matter and radiation eras, we shall only investigate the stability of this regime for $\theta=3$ and $\lambda=1$ to establish whether it is attainable in a realistic cosmological background. In Fig. \ref{ultrarelativistic}, we plot the region of parameter space $(\beta,\alpha)$ for which the ultra-relativistic linear scaling regime is excluded using both calibrations. This figure clearly shows that the values of $\beta$ for which this regime may be attained may be severely reduced for some values of $\alpha$. However, since a definite calibration of $k$ and $\cc$ does not exist for all values and $\beta$ and $\alpha$, the shape of the exclusion region cannot definitely be established. In any case, it is clear that, for the values of $\alpha$ that describe current simulations more adequately ($\alpha \gtrsim 1.15$, so that $k_m$ and $k_r$ may be reconciled), this regime is unstable in both matter and radiation-dominated epochs. Curiously, when one uses the radiation-era calibration ($k=k_r$ and $\cc=\cc_r$), this regime seems also to be excluded in the VOS model for global monopoles introduced in \cite{Martins:2008zz}.

Here, we should note that, in the region of parameter space $(\beta,\alpha)$  in which it is allowed, the ultra-relativistic linear scaling regime is the attractor solution of the VOS equations instead of the subluminal linear scaling regime discussed in the previous section. This has to be taken into account when using simulations with small $\beta$ to calibrate the VOS model. However, it is also important to stress that the VOS model is not expected to accurately describe the macroscopic dynamics of global monopole networks in the ultra-relativistic regime. First of all, strictly speaking, the expression for the force between monopoles in Eq. (\ref{forceglobal}) is only valid in the non-relativistic limit. One would expect that, as a monopole network approaches an ultra-relativistic regime, corrections to this force and to their mass (and consequently to the acceleration term) will become increasingly relevant. Moreover, in this regime, the characteristic length of the network (defined in Eq. (\ref{rhonew})) is no longer an accurate measure of the average inter-monopole physical distance, and a single lengthscale is not expected to be sufficient to describe their dynamics. Thus, in this regime, the VOS model will lose its capability to predict global monopole network evolution. In any case, one would expect the process of monopole and anti-monopole annihilation to accelerate in this ultra-relativistic limit. If several monopoles and anti-monopoles per cosmological horizon exist, one would then expect $\mathcal{N}$ to decrease rather quickly as the ultra-relativistic regime is reached. Thus, the relative importance of the acceleration caused by the interaction between monopoles with respect to the deceleration caused by Hubble damping is expected to decrease. Furthermore, in the ultra-relativistic regime, the effects of dynamical friction --- resulting from the gravitational scattering of particles in the gravitational field of the monopole (due to the existence of a deficit angle in the metric around the monopole \cite{Barriola:1989hx}) --- are expected to become relevant (as is the case for cosmic strings \cite{Garfinkle:1986nn,Avelino:1995pm}). This dynamical friction is expected to cause a transfer of part of the monopoles' momenta to the background fluid and, thus, it will be an additional source of damping in this regime. For these reasons, we do not expect the cosmological evolution of global monopoles to result in a luminal scaling regime.

Finally, we note that monopoles involved in the final stages of the process of monopole and anti-monopole annihilation are expected to become ultra-relativistic (this effect is even more important in the case of local monopoles due to the rapid increase of the magnitude of the force as the distance between monopoles decreases). Hence, some care must be taken in order to ensure that this contribution does not dominate the estimate of the RMS velocity in numerical simulations of monopoles network evolution.

\section{Local Monopoles}\label{local}

The formation of local or magnetic monopoles \cite{Dirac:1931kp} --- so named because they carry a magnetic charge --- may occur when there is a gauge symmetry breaking \cite{tHooft:1974kcl,Polyakov:1974ek}. The properties of local monopoles differ from those of global monopoles, and, thus, their dynamics cannot be described by the VOS model we have described in the previous section. In this section, we will briefly review the properties of local monopoles, and construct a VOS model to describe their dynamics.

The positions of local monopoles and anti-monopoles were found to be correlated \cite{Einhorn:1980ik}. This means that the characteristic lengthscale --- which is a measure of the average density of local monopoles --- and the correlation or persistence length --- the distance above which monopole positions are uncorrelated --- are different. This may mean that one lengthscale may be insufficient to accurately describe local monopole networks. Note however that if the correlation length scales with $L$, the construction of a one-scale model is still possible. In this section, we shall assume that this is the case, and discuss what form such a VOS model for local monopoles should take.

Unlike global monopoles, the energy of local (or magnetic) monopoles is essentially localized within a finite region and, thus, their energy density is given by Eq. (\ref{denene}) with $p=0$. This does not mean, however, that they do not interact. As a matter of fact, local monopoles carry a magnetic charge $g=4\pi/e$, and thus there is electromagnetic interaction between them (see e.g. \cite{vilenkin2000cosmic,Zichichi:1985ci}). The force between monopoles is then of the form $F\sim g^2/L^2$, and thus the acceleration felt by the monopole is

\be
a\sim \frac{g^2}{ML^2}\equiv \frac{k}{\eta L^2}\,,
\ee
in the non-relativistic limit, where we have used the fact that the monopole mass, in this case, is given by $M\sim g\eta.$

As was the case for global monopoles, a local monopole will feel the (electromagnetic) force exerted by each monopole and anti-monopole located within its causal volume. However, due to the correlations between the positions of monopoles and anti-monopoles found in \cite{Einhorn:1980ik}, the average number of monopoles/anti-monopoles in causal contact with any given monopole may not be exactly equal to the average number of monopoles per cosmological horizon $\mathcal{N}$. Note also that the magnitude of the electromagnetic force between local monopoles decreases as the distance between monopoles (squared) increases and, thus, the expectation that the total acceleration felt by a monopole or anti-monopole would be $a \mathcal{N}^{1/2}$ may not be realistic, and we may expect a weaker dependence on $\mathcal{N}$. For this reason, as for the case of global monopoles, we shall introduce a free parameter $\alpha$. Moreover, the authors of \cite{Martins:2008zz} claim that, for global monopoles, the average number of local monopoles and anti-monopoles per cosmological horizon is given by

\be
\mathcal{N}_l=\left(\frac{d_H}{L}\right)^2\,.
\ee
Note however that this claim is inconsistent with the definition of the characteristic lengthscale and there is no evidence to support it. As a matter of fact, it follows from the definition of $L$ that $\mathcal{N}_l$ should be given by Eq. (\ref{number}). We, then, include an acceleration term of the form

\be
\frac{k}{\eta L^2}\left(\frac{d_H}{L}\right)^{\alpha}\,,\label{curvlocal}
\ee
in the evolution equation for $\vv$. Again, we opt to leave $\alpha$ as a free parameter, given the discussion we had previously in this section. However, we shall note that, as for global monopoles (and for the same reason), we expect $\alpha\ge 0$. Since the magnitude of the force between monopoles decreases with the distance, the total acceleration felt by a monopole should be mainly determined by the force exerted by the closest monopole or anti-monopole and, thus, we shall expect $\alpha$ to be smaller than that of global monopoles. This is, however, a complex problem that can only be fully addressed with numerical simulations.

It has been demonstrated \cite{Zeldovich:1978wj,Preskill:1979zi,Martins:2008zz} that, given the electromagnetic nature of monopole interactions, the form of the energy loss term (caused by monopole and anti-monopole annihilations) should be different from that of global monopoles:

\be
\left.\dtot{L}{t}\right|_{ann.}=\frac{C}{3}\frac{\eta^{p-2}}{L^2 T^p}\,, \label{losslocal}
\ee
where $C$ is an adimensional constant, $T$ is the background temperature and one should expect $p\le 3$ \cite{Preskill:1979zi}. In Ref. \cite{Preskill:1979zi}, it was suggested that there should be a high temperature transient regime during which $p=2$ followed by a regime in which $p=9/10$.

Given Eqs. (\ref{curvlocal}) and (\ref{losslocal}), the VOS model for local monopoles should take the form

\bq
\dtot{\vv}{t} & = & (1-\vv^2)\left[\frac{k}{\eta L^2}\left(\frac{d_H}{L}\right)^{\alpha}-\frac{\vv}{\ell_d}\right]\,,\label{vosvlocal}\\
\dtot{L}{t} & = & HL+\frac{\vv^2}{3} \frac{L}{\ell_d}+C\frac{\eta^{p-2}}{3L^2 T}\,,\label{vosLlocal}
\eq
where $\ell_d^{-1}=H+\ell_f^{-1}$ and $\ell_f$ is the frictional lengthscale which, given the nature of local monopoles, should be determined by their interactions with charged particles.

This VOS model for local monopoles differs from that of Ref. \cite{Martins:2008zz} only in the form of the acceleration term. However, this change makes a significant difference in the type of evolution these equations allow for the dynamics of local monopoles. In particular, in \cite{Martins:2008zz}, the authors found that their model predicts the existence of scaling regimes with

\be
L\propto a\,,\quad\mbox{and}\quad\vv\propto a^{-1}\,,
\ee
for $p<3-1/\beta$, and

\be
L\propto t^{(\beta p+1)/3}\,,\quad\mbox{and}\quad\vv\propto a^{-1}\,,
\ee
for $p>3-1/\beta$. It is straightforward to show that these regimes can only arise for $\alpha=-1$ --- which corresponds to the model they have built --- and, since $\alpha\ge 0$ in our physically motivated model, these regimes are not expected to exist. On the other hand, Eqs. (\ref{vosvlocal}) and (\ref{vosLlocal}) admit an ultra-relativistic regime of the form

\be
L\propto a^Y\,,\quad\mbox{and}\quad\vv=1\,,
\ee
for $p<2$ (corresponding to the case in which the effect of the energy loss on the dynamics is negligible) and $p=2+Y-1/\beta$, with $Y\le 1+1/\beta$ (which would arise when the energy loss process is the main factor affecting the dynamics of the network). However, as we have discussed in the previous section, we do not expect the predictions of the VOS model to hold in the ultra-relativistic regime. As a matter of fact, as discussed for global monopoles, one shall not expect luminal scaling regimes to be stable on physical grounds: the relative importance of the Hubble damping term, when compared to the acceleration term, is expected to increase as the network approaches these regimes and the effects of dynamical friction are expected to become relevant.

\section{Conclusions}\label{conc}

In this paper, we have revisited the VOS model for global and local monopoles, with particular emphasis on the global case. We have proposed a physically motivated change to the acceleration term of the RMS velocity equation of motion and have demonstrated that such a change is necessary in order for the VOS model to accurately describe the most recent numerical simulations of global monopole networks with a unique value for the acceleration parameter $k$. However, we have also shown that, although this change is necessary, it is not sufficient for an accurate description of their dynamics. The fact that the energy of a global monopoles is not localized within its core complicates the problem significantly. Although this means that the coefficients of the Hubble damping terms should be different from those of point particles, it is not clear for now which value these coefficients should take. Moreover, given the complexity of this problem, it is not clear whether a simple velocity-dependent one-scale model is sufficient to describe the intricacy of global monopole dynamics. Settling this question will require further analytical and numerical modeling of global monopole network evolution.

Nevertheless, we shall note that, despite these open questions, the changes to the VOS model for global monopoles we have proposed --- both to the acceleration term and to the Hubble damping coefficients --- already allow for a more adequate description of numerical simulations by resolving the tensions between the radiation and matter era simulation results (since with this changed model one may find a unique calibration of $k$ and $\cc$ that fits both simulations). We have also proposed the corresponding changes to the acceleration term of the VOS model for local monopoles. In this case, no simulations to assert the validity of these changes exist. However, on physical grounds, we shall expect this new form of the acceleration term to also provide a more adequate description of realistic local monopole networks.

This work also has implications for analytical studies of the dynamics of hybrid defect networks in which monopoles are connected by cosmic strings \cite{Martins:2009hj,Nunes:2011sf,Achucarro:2013mga}. The case of semi-local strings --- which are (non-topological) open-ended strings whose ends behave as global monopoles --- is of particular physical relevance since their production is predicted in some brane inflationary models \cite{Urrestilla:2004eh,Dasgupta:2004dw}. Some attention was, thus, naturally devoted to numerical simulations of their dynamics and to developing VOS-type models to describe these simulations. These models have been based on the VOS model for monopoles introduced in \cite{Martins:2008zz} --- which has, as we have pointed out, some flaws --- and thus they should not be expected to provide an accurate depiction of the evolution of these networks. We expect that using our VOS model for global and local monopoles will also improve the analytical description of hybrid defect networks and of networks of semi-local strings.

\acknowledgments
L.S. is supported by Funda\c{c}\~{a}o para a Ci\^{e}ncia e Tecnologia (FCT, Portugal) and by the European Social Fund (POPH/FSE) through the grant SFRH/BPD/76324/2011. Funding of this work was also provided by the FCT grant UID/FIS/04434/2013.

\bibliography{monopoles}

\end{document}